\documentclass[11pt]{article}
\usepackage{epsf}
\usepackage{amsmath}
\usepackage{graphics}
\usepackage{cite}

\renewcommand{\thefootnote}{\#\arabic{footnote}}
\begin{document}

\newcommand{\gtrsim}{ \mathop{}_{\textstyle \sim}^{\textstyle >} }
\newcommand{\lesssim}{ \mathop{}_{\textstyle \sim}^{\textstyle <} }

\newcommand{\rem}[1]{{\bf #1}}

\renewcommand{\thefootnote}{\fnsymbol{footnote}}
\setcounter{footnote}{0}
\begin{titlepage}

\def\thefootnote{\fnsymbol{footnote}}

\hfill September 2017\\
\vskip .5in
\bigskip
\bigskip

\begin{center}
{\Large \bf  Holographic Entanglement Entropy and Cyclic Cosmology}

\vskip .45in

{\bf Paul H. Frampton\footnote{email: paul.h.frampton@gmail.com}}

{Dipartimento di Matematica e Fisica "Ennio De Giorgi"\\
 Universit\`a del Salento and INFN Lecce, 
 Via Arnesano 73100 Lecce,Italy.}

\end{center}

\vskip .4in
\begin{abstract}
\noindent
We discuss a cyclic cosmology in which the visible universe, or introverse, is all
that is accessible to an observer while the extroverse represents the total
spacetime originating from the time when the dark energy began to
dominate. It is argued that entanglement entropy of the introverse is the
more appropriate quantity to render infinitely cyclic, rather than the 
entropy of the total universe. Since vanishing entanglement entropy
implies disconnected spacetimes,  at the turnaround when the introverse
entropy is zero the disconnected extroverse can be jettisoned
with impunity.
\end{abstract}

\end{titlepage}

\renewcommand{\thepage}{\arabic{page}}
\setcounter{page}{1}
\renewcommand{\thefootnote}{\#\arabic{footnote}}

\newpage

\section{Introduction}

\noindent
Cyclic cosmology with an infinite number of cycles which include -- expansion,
turnaround, contraction, bounce -- is an appealing
theory, and one of the most recent attempts at a solution was in \cite{BF}.
As emphasized first by Tolman\cite{Tolman1,Tolman2}, entropy and the second law of thermodynamics
provide a significant challenge to implementing a cyclic theory.

\bigskip

\noindent
The approach was based on the so-called Come Back Empty (CBE) principle
which chooses the turnaround as the point in the cycles where entropy
is jettisoned. A successful universe must be selected with zero entropy
at the turnaround to begin an adiabatic contraction to the bounce
while empty of all matter including black holes.
In this article we provide a better understanding of CBE based on 
the holographic principle\cite{holo,Susskind,Maldacena}. In particular, the notion of
entanglement entropy\cite{RT} and the related
question of spacetime connectivity\cite{Raamsdonk1,Raamsdonk2}
can put the CBE model on a quantum theoretical basis.

\bigskip

\noindent
Although the holographic principle was well-known
at that time, its relationship to cyclic cosmology was unclear. Even how it could
help reconcile general relativity with quantum mechanics was less developed.
In the AdS/CFT correspondence, it has been argued that the dynamics of spacetime
on the AdS side, including the connectivity of spacetime,
is related to quantum entanglement of
disconnected classical theories on the CFT side. This
idea was introduced by Van Raamsdonk in an essay submitted to the Gravity
Research Foundation. Considerations of the entropy
of the universe have been discussed also in \cite{FHKR} and by
E. Verlinde\cite{Verlinde1,Verlinde2}

\bigskip

\noindent
Before discussing the impact on cyclic cosmology which is our 
present subject, let us briefly discuss how this line of reasoning
changes the marriage of general relativity with quantum mechanics.
For several decades it had been tacitly assumed that there exists
a quantum gravity theory whose classical limit 
is general relativity. Now it appears that whether or not
such a theory exists may not be the best question to ask. Classical
general relativity which describes the dynamics of spacetime may
instead arise from consistency conditions on the quantum entanglement of
conformal field theories which are at the holographic boundary. 

\bigskip

\noindent
Subsequent work has been mainly on time-independent spacetimes. 
However, we shall show that for the
the time-dependent CBE model of cyclic cosmology use of holographic
quantum entanglement entropy makes possible a better
understanding of entropy at the turnaround.

\bigskip

\noindent
First we briefly review the CBE model.

\newpage

\section{Extroverse and Introverse in Expansion Era}

\noindent
The best way to discuss the CBE model is by using the language
introduced in \cite{PHF2} where the spacetime occupied by the
universe is divided into two subregions denoted by introverse
and extroverse respectively which we must define carefully because
it will be the entanglement entropy between these two subregions
which will play a central r\^ole in the cyclicity.

\bigskip

\noindent
The present visible universe at $t=t_0$  is the present introverse and
its comoving radius is
\begin{equation}
R_{int}(t_0) = 44 Gly
\label{VUradius}
\end{equation}
A more useful time to consider than the present $t_0 = 13.8$Gy is the time $t=t_{DE}$
when the dark energy began to dominate over the matter component. This occurred
at $t_{DE}=9.8$Gy at which time the extroverse is identified with the introverse.
After $t=t_{DE}$, the extroverse expands exponentially and become larger, eventually
very much larger, than the introverse. 

\bigskip

\noindent
At $t=t_{DE}$
\begin{equation}
R_{ext}(t_{DE}) = R_{int}(t_{DE}) = 39Gly
\label{tDE}
\end{equation}
at a time when the scale factor, normalized to $a(t_0)=1$, was $a(t_{DE})=0.75$.

\bigskip

\noindent
The extroverse expands exponentially for $t_{DE} < t < t_T$, where $t_T$ is the turnaround
time calculated to be $t_T = 1.3$Ty in \cite{PHF3}. The extroverse comoving radius during this
expansion is given by
\begin{equation}
R_{ext}(t) = \left( 39 Gly \right) {\rm exp} \left[ \frac{t-9.8Gy}{13.8Gy} \right]
\label{Rext}
\end{equation}
Two interesting values at $t=t_0$ and $t=t_T$ respectively are
\begin{equation}
R_{ext}(t_0) = 52 Gly ~~~~~ {\rm and} ~~~~~ R_{ext}(t_T) = 1.5 \times 10^{42} Gly
\label{Rext}
\end{equation}

\bigskip

\noindent
The introverse expands more gradually for $t_{DE} < t < t_T$, being limited by the speed
of light in its definition. The introverse comoving radius for the expansion era is given in terms
of $a(t) = {\rm exp} [ (t - 13.8 Gy)/(13.8 Gy)]$ by
\begin{eqnarray}
R_{int}(t) & = & 39 Gly + c \int_{t_{DE}}^{t} dt a(t)^{-1}   \nonumber  \\
&= & 39 Gly + \left( 13.8 Gly \right) \left[ 1 - {\rm exp} \{- (t - 9.8Gy) / (13.8Gy)\} \right] \nonumber \\
& \equiv & 44Gly +  \left( 13.8 Gly \right) \left[ 1 - {\rm exp} \{- (t - 13.8Gy) / (13.8Gy)\} \right]  
\label{Rinverse}
\end{eqnarray}
   
\bigskip
   
\noindent
From Eq.(\ref{Rinverse}) the values of $R_{inv}(t)$ at $t=t_0$ and $t=t_T$ are
\begin{equation}
R_{int}(t_0) = 44 Gly ~~~~~ {\rm and} ~~~~~ R_{inv}(t_T) = 58 Gly
\label{Rint}
\end{equation} 

\bigskip

\noindent
The behaviour of $R_{ext}(t)$ and $R_{inv}(t)$ calculated in (\ref{Rext}) and (\ref{Rint})
respectively assumes that the dark energy is accurately described by a cosmological constant.  
The results for $R_{ext}(t_0)$ and $R_{inv}(t_0)$ show that the present extroverse is 
already $60\%$ larger in volume
than the introverse implying that hundreds of billions of galaxies have exited the introverse
since the onset of exponential expansion.

\newpage

\section{The Turnaround}

\noindent
In the previous section we used the value $t_T = 1.3$Ty without
further comment. It is quite simple to re-derive that this must be the turnaround 
time in the CBE model as follows.

\bigskip

\noindent
For infinite cyclicity, it is necessary to impose a matching condition on the
contraction scale factor $\hat{a}(t)$ defined in \cite{BF} as
\begin{equation}
\hat{a}(t) \equiv \left[ \frac{R_{inv}(t_T)}{R_{ext}(t_T)} \right] a(t)
\label{ahat}
\end{equation}
with the scale factor $a(t)$ of the previous expansion. Since the contraction
has a radiation-dominated behaviour $\sim \sqrt{t}$ the matching
is conveniently made at the onset of matter domination during expansion
when $t=t_m=47ky$ and the expansion scale factor $a(t_m)$ is
\begin{equation}
a(t_m) = \left( 0.75 \right) \left[ \frac{47ky}{9.8Gy} \right]^{\frac{2}{3}} = 2.1\times10^{-4}
\label{am}
\end{equation}

\bigskip
 
\noindent
The value of $\hat{a}(t_T)$ follow from Eq.(\ref{ahat})  to be
\begin{equation}
\hat{a}(t_T) = \left( \frac{R_{inv}(t_T)}{a(t_T) R_{ext}(t_0)} \right) a(t_T) = \left( \frac{R_{inv}(t_T)}{R_{ext}(t_0)} \right) = 1.11.
\label{ahatT}
\end{equation}
The infinite-cyclicity matching condition
\begin{equation}
\hat{a}(t_m) \equiv a(t_m)
\label{matching}
\end{equation}
then provides the unique solution
\begin{equation}
t_T = (47ky) \left( \frac{1.11}{2.1\times 10^{-4}} \right)^{\frac{1}{2}} = 1.3Ty
\label{tT}
\end{equation}
for the turnaround time. Note that this could not be derived in \cite{BF} because
there the equation of state for the dark energy was assumed to be $\omega <-1$
but cyclicity is now known to require\cite{PHF2}  $\omega=-1$ precisely.

\bigskip

\noindent
We shall improve the original version of Comes Back Empty (CBE) 
with a new understanding based on quantum entanglement. Although
quantum fluctuations as precursors of cosmic structure formation
are well-known {\it e.g.} \cite{PHFfluctuations}, to our
knowledge this is the first application of quantum computation and quantum
information \cite{NielsenChuang}, in particular quantum entanglement, to
theoretical cosmology. To see that it is an improvement, let us first describe
the original CBE model.

\bigskip

\noindent
In the original derivation which used phantom dark energy with $\omega < -1$,
a modified Friedmann equation in the form
\begin{equation}
\left( \frac{\dot{a}}{a} \right)^2 = \frac{8 \pi G} {3} \left[ \rho - \frac{\rho^2}{\rho_{cr}} \right]
\label{modified}
\end{equation}
was used where $\rho_{cr}$ is a constant. Eq.(\ref{modified}) can be derived from string theory.
This led to a bounce when $\rho = \rho_{cr}$
at a time $< 10^{-27}$s before the Big Rip. The scale factor deflated dramatically to
$\hat{a}(t_T) = f a(t)$ where $f < 10^{-28}$. All bound states become unbound and soon 
afterwards become causally disconnected. The critical density is $\rho_{cr} = \eta \rho_{H_2O}$
with $\eta$ between $10^{31}$ and $10^{87}$, safely below the Planck value
$\eta = 10^{104}$. A given patch is chosen so that it is empty of all matter including
black holes, and contains only dark energy. This causal patch contracts
adiabatically to a bounce with zero entropy.

\bigskip

\noindent
In the present derivation, the dark energy is described by a cosmological constant 
with $\omega = -1$ so that
Eq.(\ref{modified}) does not turnaround since $\rho(t)$ now asymptotes to a constant. Therefore
we must use the unmodified Friedmann equation
\begin{equation}
\left( \frac{\dot{a}}{a} \right)^2 = \frac{8 \pi G \rho} {3}  
\label{unmodified}
\end{equation}
at almost all times except {\it times extremely close to the turnaround or the bounce}. Since this involves
modification of general relativity we take a phenomenological approach with
\begin{equation}
\left( \frac{\dot{a}}{a} \right)^2 = \frac{8 \pi G \rho} {3} 
\left[ \left( 1 - \frac{a(t)^2}{a(t_T)^2} \right) \left( 1 - \frac{a(t_B)^2}{a(t)^2} \right) \right]
\label{remodified}
\end{equation}
where $t_T, t_B$ are the turnaround, bounce times respectively. In Eq.(\ref{remodified}), the
square bracket equals one at almost every time.
We shall not employ
Eq.(\ref{remodified}) in the following and it is written only as an illustration.

\bigskip

\noindent
Now we turn to the r\^ole of quantum entanglement. This begins from the holographic
principle and its realization in the AdS/CFT duality.
One previous use of quantum mechanics in cosmology has been the idea that
quantum fluctuations of the inflaton field during inflation are amplified
to seed large scale structure formation. Here we suggest that
quantum entanglement and its relationship with spacetime connectivity
play an comparably dramatic r\^ole by elucidating the CBE
cyclic cosmology model at turnaround.

\bigskip

\noindent
In the AdS/CFT correspondence, if we consider two 
non-interacting identical copies $CFT_A$ and $CFT_B$, a state of the
system can in general be written
\begin{equation}
|\Psi> = |\Psi>_A \otimes |\Psi>_B.
\label{product}
\end{equation}
The CFTs are each dual to separate asymptotically AdS spacetimes
so that the direct product (\ref{product}) is dual to two spacetimes which
are disconnected.

\bigskip

\noindent
As a quantum state we are free to consider a superposition of states.
Let the energy eigenstates be $|E_k>$ for one CFT and consider the 
quantum superposition
\begin{equation}
|\Psi (\beta) > = \Sigma_k e^{-\frac{\beta}{2}E_k} |E_k>\otimes|E_k>
\label{superposition}
\end{equation}
which can be shown, in general, to be dual to a single connected spacetime.
Starting with $|\Psi>$ we deduce that the denstiy matrix for $CFT_B$ is
\begin{equation}
Tr \left( |\Psi> <\Psi| \right) = \Sigma_k e^{-\beta E_k} |E_k><E_k| = \rho_T
\label{thermal}
\end{equation}
which is a thermal density matrix. so the quantum superposition of two
disconnected spacetimes is identified with a classical connected spacetime.
This leads to the fascinating idea that 
{\it classical spacetime connectivity arises by quantum entangling the degrees of freedom in two components}
which is a central idea in what follows.

\bigskip

\noindent
Proceeding in the opposite direction, let us consider one CFT on a sphere and divide the sphere
initially into two hemispheres A and B with Hilbert space ${\cal H}_A\otimes{\cal H}_B$.
According to \cite{NielsenChuang} the entanglement entropy is the von Neumann entropy
\begin{equation}
S(A) = - Tr \left( \rho_A {\rm log} \rho_A \right)
\label{vonNeumann}
\end{equation}
with
\begin{equation}
\rho_A = Tr_B \left( |\Psi><\Psi| \right).
\label{rhoA}
\end{equation}
 As $S(A)$ decreases, the area of the minimum surface separating A and B decreases
 according to the Ryu-Takayanagi prescription\cite{RT} which greatly generalises
 the original Bekenstein-Hawking area formula for the entropy of a black hole.
 The sphere becomes
 a dumbbell shape. In the limit $S(A) \rightarrow 0$, the spacetimes A and B become 
 classically disconnected.

\bigskip

\noindent
Let us briefly consider entanglement of two-state quantum mechanical systems,
or quantum bits usually shortened to qubits. One qubit can be in the general state
\begin{equation}
\Psi = \alpha_{\uparrow} |\uparrow> + \alpha_{\downarrow}|\downarrow>
\label{onequbit}
\end{equation}
with probabilities $|\alpha_{\uparrow}|^2$ and $|\alpha_{\downarrow}|^2$ of measuring $|\uparrow>$ and
$|\downarrow>$ respectively.

\bigskip

\noindent
For two qubits the most general state is
\begin{equation}
|\Psi> = \alpha_{\uparrow\uparrow}|\uparrow\uparrow> + \alpha_{\uparrow\downarrow}|\uparrow\downarrow> +
\alpha_{\downarrow\uparrow}|\downarrow\uparrow> +
\alpha_{\downarrow\downarrow}|\downarrow\downarrow> 
\label{twoqubits}
\end{equation}
with $\Sigma |\alpha_{ij}|^2 = 1$.

\bigskip

\noindent
If we measure the first qubit to be $|\uparrow>$, then the normalized post-measurement state
is
\begin{equation}
|\Psi>^{'} = \left[ \frac{\alpha_{\uparrow\uparrow} |\uparrow\uparrow> + 
\alpha_{\uparrow\downarrow} |\uparrow\downarrow>}
{\sqrt{|\alpha_{\uparrow\uparrow}|^2 + |\alpha_{\uparrow\downarrow}|^2}} \right]
\label{postmeasure}
\end{equation}
The phenomenon of entanglement is clearest in an EPR state
\begin{equation}
|\Psi> = \frac{1}{\sqrt{2}} \left[ |\uparrow\uparrow> + |\downarrow\downarrow> \right]
\label{EPR}
\end{equation}
named for a paper\cite{EPR} which tried unsuccessfully to prove that
quantum mechanics is incomplete. Eq.(\ref{EPR}) is sometimes alternatively called a Bell
state\cite{Bell} for the physicist who derived inequalities based on local realism
which are violated by quantum mechanics.
The correlations which occur in quantum mechanics are stronger than
could ever exist between classical systems. The point 
about the EPR/Bell state Eq.(\ref{EPR}) is that measuring the first qubit as $|\uparrow>$
implies that measurement of the second qubit must with 100\% certainty result 
in $|\uparrow>$, a correlation which never happens in classical mechanics.

\bigskip

\noindent
After these digressions, we return to our central point of the turnaround in cyclic cosmology.
Alternative approaches to cyclic cosmology include \cite{Steinhardt} and \cite{Penrose} .

\newpage

\noindent
The CBE model with its empty contraction era aims to jettison 
the accrued entropy at the turnaround so that the contracting
universe has zero entropy. One reason for this is that the presence of matter including black holes
will disallow contraction to the bounce because black holes will merge and enlarge and 
in the presence of any matter going in reverse
through phase transitions will be entropically impossible. Thus the idea was that the causal
patch at turnaround which is selected from a very large number of candidates, as suggested
by the comparison between the two turnaround radii in Eqs.(\ref{Rext})
and (\ref{Rint}), will almost always be empty of matter. The tiny fraction which do contain matter
will represent failed universes which will experience a premature bounce rather than a
successful contraction.

\bigskip

\noindent
Mathematicians or string theorists may provide details of a non-singular turnaround
but until then we remain temporarily satisfied with the phenomenological Eq.(\ref{remodified})
with a view first to discover a scenario which {\it can} be infinitely cyclic.

\bigskip

\noindent
As the extroverse spacetime is stretched by the extreme amount in Eq.(\ref{Rext})
it will become disconnected into causal patches and this disconnection
corresponds precisely to making $S(int)$ vanish where $S(int)$ is the
entanglement entropy between introverse and extroverse.
Because the extroverse becomes a classically disconnected spacetime, we may
study the contracting introverse without any further concern for the extroverse
whose entropy is thereby jettisoned. 

\bigskip

\noindent
Tolman's work had so much impact that almost
no work on cyclic cosmology was pursued thereafter. The most important subsequent step was the discovery
of dark energy in 1998 which provides the clear division into introverse and extroverse as discussed
in the previous section. Tolman was necessarily considering the entropy
for the whole universe and applied the second law of thermodynamics to it, to arrive at
his impossibility theorem.

\bigskip

\noindent
Tolman was certainly aware of quantum mechanics but it surely never entered his head
that quantum mechanics is relevant to cosmology. Such an idea only emerged long after
Tolman died in 1948, in the 1980s when quantum fluctuations were proposed to seed
structure formation. Now we
learn that, at the other end of the expansion era, quantum mechanics may play
a r\^ole also in the spacetime disconnection of the introverse, as well as 
in choosing the quantum entanglement entropy of the introverse to be
the quantity which is infinitely cyclic.

\bigskip

\noindent
As for experimental tests of CBE cyclic cosmology, regrettably only one is immediately
apparent which is the
prediction that the equation of state of dark energy is a constant
$\omega = -1$ very precisely to many decimal places. If a measurable departure
from this prediction were found experimentally, it would refute the CBE 
cyclic model. Hopefully with more work on the model, other testable
predictions will be discovered.

\newpage

\section{Discussion}

\noindent
Quantum mechanics was originally discovered as a description for 
atomic physics with no anticipation that it would be useful in theoretical
cosmology. The first such use was several decades later based on quantum
perturbations in the early universe seeding the growth of large scale
structure; this is well-known in inflationary scenarios and less known
\cite{PHFfluctuations} in cyclic bounce models.

\bigskip

\noindent
In the present article we have suggested a second 
use of quantum mechanics in theoretical cosmology.
It arose unexpectedly by better understanding the CBE model\cite{BF}.
The underlying quantum theory is based on AdS/CFT duality 
and entropic entropy where an important source of encouragement is the emergent understanding 
(see {\it e.g.} \cite{Preskill}) of AdS/CFT and the Ryu-Takayanagi formula
from analysis of holographic quantum error-correcting codes.

\bigskip

\noindent
It is a decade since publication of \cite{BF} which did generate a lot of discussion at that time
\footnote{Mostly because an infinite past seemed to many to contradict their religious view of creation.}.
In addressing Tolman's no-go theorem, the present improved suggestion is to make
an infinitely-cyclic cosmology by requiring infinite cyclicity not of the
entropy of the whole universe but only of the quantum entanglement entropy
of the introverse, $S(int)$.

\bigskip

\noindent
This has the advantage that when $S(int)=0$
at turnaround, the introverse becomes a spacetime manifold
disconnected from the extroverse and hence the latter
can be jettisoned with impunity. This avoids the appearance
of an infiniverse after infinitely many cycles
and enables a model with only one universe. 

\vspace{1.0in}

\noindent
\section*{Acknowledgement}

\noindent
We thank INFN for support and the Department of Physics
at the University of Salento for hospitality.

\end{document}